\documentclass[manyauthors,nocleardouble,COMPASS]{cernphprep}

\usepackage{graphicx}
\usepackage{subfig}
\usepackage[percent]{overpic}
\usepackage[numbers, square, comma, sort&compress]{natbib}

\usepackage{times}
\usepackage{epsfig}
\usepackage{amssymb}
\usepackage{colordvi}
\usepackage{graphicx}
\usepackage{wrapfig,rotating}
\usepackage{amsmath}
\usepackage{hyperref} 
\usepackage{caption}

\usepackage{lineno}
\usepackage{dcolumn}

\begin {document}

\begin{titlepage}
\PHnumber{2012--189}
\PHdate{03 July 2012}
\title{Measurement of the Cross Section for High-$p_T$ Hadron Production in Scattering of 160\,GeV/$c$ Muons off Nucleons}

\Collaboration{The COMPASS Collaboration}
\ShortAuthor{The COMPASS Collaboration}

\begin{abstract}
  The differential cross section for production of charged hadrons with high transverse
  momenta in scattering of 160\,GeV/$c$ muons off nucleons at low photon
  virtualities has been measured at the COMPASS experiment at CERN. The results,
  which cover transverse momenta from 1.1\,GeV/$c$ to 3.6\,GeV/$c$, 
  are compared to a  
  perturbative Quantum Chromodynamics (pQCD)
  calculation, 
  in order to evaluate the applicability of pQCD to this process in
  the kinematic domain of the experiment. The shape of the calculated
  differential cross section as a function of transverse momentum is found to be
  in good agreement with the experimental data, but the 
  absolute scale is 
  underestimated by next-to-leading order (NLO) pQCD. The inclusion of 
  all-order resummation of large logarithmic threshold corrections 
  reduces
  the discrepancy from a factor of three to four to a factor of two. 
  The dependence of the cross section
  on the pseudo-rapidity and on virtual photon energy fraction is investigated. 
  Finally the dependence on the charge of the hadrons is discussed.
\end{abstract}

\vfill \Submitted{(to be submitted to Phys.~Rev.~D)}
\end{titlepage}

{\pagestyle{empty} 
%
%

\section*{The COMPASS Collaboration}
\label{app:collab}

\begin{flushleft}
C.~Adolph\Irefn{erlangen},
M.G.~Alekseev\Irefn{triest_i},
V.Yu.~Alexakhin\Irefn{dubna},
Yu.~Alexandrov\Irefn{moscowlpi}\Deceased,
G.D.~Alexeev\Irefn{dubna},
A.~Amoroso\Irefn{turin_u},
V.~Andrieux\Irefn{saclay},
A.~Austregesilo\Irefnn{cern}{munichtu},
B.~Bade{\l}ek\Irefn{warsawu},
F.~Balestra\Irefn{turin_u}
% ...
\end{flushleft}

%
%

\begin{Authlist}
\item \Idef{bielefeld}{Universit\"at Bielefeld, Fakult\"at f\"ur Physik, 33501 Bielefeld, Germany\Arefs{f}}
\item \Idef{bochum}{Universit\"at Bochum, Institut f\"ur Experimentalphysik, 44780 Bochum, Germany\Arefs{f}}
\item \Idef{bonniskp}{Universit\"at Bonn, Helmholtz-Institut f\"ur  Strahlen- und Kernphysik, 53115 Bonn, Germany\Arefs{f}}
\item \Idef{bonnpi}{Universit\"at Bonn, Physikalisches Institut, 53115 Bonn, Germany\Arefs{f}}
\item \Idef{brno}{Institute of Scientific Instruments, AS CR, 61264 Brno, Czech Republic\Arefs{g}}
\item \Idef{calcutta}{Matrivani Institute of Experimental Research \& Education, Calcutta-700 030, India\Arefs{h}}
\item \Idef{dubna}{Joint Institute for Nuclear Research, 141980 Dubna, Moscow region, Russia\Arefs{i}}
\item \Idef{erlangen}{Universit\"at Erlangen--N\"urnberg, Physikalisches Institut, 91054 Erlangen, Germany\Arefs{f}}
\item \Idef{freiburg}{Universit\"at Freiburg, Physikalisches Institut, 79104 Freiburg, Germany\Arefs{f}\Arefs{ll}}
\item \Idef{cern}{CERN, 1211 Geneva 23, Switzerland}
\item \Idef{liberec}{Technical University in Liberec, 46117 Liberec, Czech Republic\Arefs{g}}
\item \Idef{lisbon}{LIP, 1000-149 Lisbon, Portugal\Arefs{j}}
\item \Idef{mainz}{Universit\"at Mainz, Institut f\"ur Kernphysik, 55099 Mainz, Germany\Arefs{f}}
\item \Idef{miyazaki}{University of Miyazaki, Miyazaki 889-2192, Japan\Arefs{k}}
\item \Idef{moscowlpi}{Lebedev Physical Institute, 119991 Moscow, Russia}
\item \Idef{munichlmu}{Ludwig-Maximilians-Universit\"at M\"unchen, Department f\"ur Physik, 80799 Munich, Germany\Arefs{f}\Arefs{l}}
\item \Idef{munichtu}{Technische Universit\"at M\"unchen, Physik Department, 85748 Garching, Germany\Arefs{f}\Arefs{l}}
\item \Idef{nagoya}{Nagoya University, 464 Nagoya, Japan\Arefs{k}}
\item \Idef{praguecu}{Charles University in Prague, Faculty of Mathematics and Physics, 18000 Prague, Czech Republic\Arefs{g}}
\item \Idef{praguectu}{Czech Technical University in Prague, 16636 Prague, Czech Republic\Arefs{g}}
\item \Idef{protvino}{State Research Center of the Russian Federation, Institute for High Energy Physics, 142281 Protvino, Russia}
\item \Idef{saclay}{CEA IRFU/SPhN Saclay, 91191 Gif-sur-Yvette, France\Arefs{ll}}
\item \Idef{telaviv}{Tel Aviv University, School of Physics and Astronomy, 69978 Tel Aviv, Israel\Arefs{m}}
\item \Idef{triest_i}{Trieste Section of INFN, 34127 Trieste, Italy}
\item \Idef{triest}{University of Trieste, Department of Physics and Trieste Section of INFN, 34127 Trieste, Italy}
\item \Idef{triestictp}{Abdus Salam ICTP and Trieste Section of INFN, 34127 Trieste, Italy}
\item \Idef{turin_u}{University of Turin, Department of Physics and Torino Section of INFN, 10125 Turin, Italy}
\item \Idef{turin_i}{Torino Section of INFN, 10125 Turin, Italy}
\item \Idef{turin_p}{University of Eastern Piedmont, 15100 Alessandria,  and Torino Section of INFN, 10125 Turin, Italy}
\item \Idef{warsaw}{National Centre for Nuclear Research, 00-681 Warsaw, Poland\Arefs{n} }
\item \Idef{warsawu}{University of Warsaw, Faculty of Physics, 00-681 Warsaw, Poland\Arefs{n} }
\item \Idef{warsawtu}{Warsaw University of Technology, Institute of Radioelectronics, 00-665 Warsaw, Poland\Arefs{n} }
\item \Idef{yamagata}{Yamagata University, Yamagata, 992-8510 Japan\Arefs{k} }
\end{Authlist}
%
%
\vspace*{-\baselineskip}
\begin{Authlist}
\item \Adef{a}{Also at IST, Universidade T\'ecnica de Lisboa, Lisbon, Portugal}
\item \Adef{aa}{Also at Department of Physics, Pusan National University, Busan 609-735, Republic of Korea}
\item \Adef{bb}{Supported by the DFG Research Training Group Programme 1102  ``Physics at Hadron Accelerators''}
\item \Adef{b}{Also at Chubu University, Kasugai, Aichi, 487-8501 Japan\Arefs{k}}
\item \Adef{c}{Also at KEK, 1-1 Oho, Tsukuba, Ibaraki, 305-0801 Japan}
\item \Adef{cc}{Also at Moscow Institute of Physics and Technology, Moscow Region, 141700, Russia}
\item \Adef{y}{present address: National Science Foundation, 4201 Wilson Boulevard, Arlington, VA 22230, United States}
\item \Adef{x}{present address: RWTH Aachen University, III. Physikalisches Institut, 52056 Aachen, Germany}
\item \Adef{e}{Also at GSI mbH, Planckstr.\ 1, D-64291 Darmstadt, Germany}
\item \Adef{f}{Supported by the German Bundesministerium f\"ur Bildung und Forschung}
\item \Adef{g}{Supported by Czech Republic MEYS Grants ME492 and LA242}
\item \Adef{h}{Supported by SAIL (CSR), Govt.\ of India}
\item \Adef{i}{Supported by CERN-RFBR Grants 08-02-91009 and 12-02-91500}
\item \Adef{j}{\raggedright Supported by the Portuguese FCT - Funda\c{c}\~{a}o para a Ci\^{e}ncia e Tecnologia, COMPETE and QREN, Grants CERN/FP/109323/2009, CERN/FP/116376/2010 and CERN/FP/123600/2011}
\item \Adef{k}{Supported by the MEXT and the JSPS under the Grants No.18002006, No.20540299 and No.18540281; Daiko Foundation and Yamada Foundation}
\item \Adef{l}{Supported by the DFG cluster of excellence `Origin and Structure of the Universe' (www.universe-cluster.de)}
\item \Adef{ll}{Supported by EU FP7 (HadronPhysics3, Grant Agreement number 283286)}
\item \Adef{m}{Supported by the Israel Science Foundation, founded by the Israel Academy of Sciences and Humanities}
\item \Adef{n}{Supported by the Polish NCN Grant DEC-2011/01/M/ST2/02350}
\item [{\makebox[2mm][l]{\textsuperscript{*}}}] Deceased
\end{Authlist}

\clearpage
}

\newpage
%
%

\section{Introduction}
\label{sec:intro}
Most of the current knowledge about the structure of the nucleon has been
derived from high-energy lepton-nucleon scattering experiments
(see e.g.\ Ref.~\cite{Thomas:2001kw}). 
The theoretical framework for the interpretation of data
from such experiments is perturbative Quantum Chromodynamics (pQCD). In the
presence of a large momentum transfer in the reaction, pQCD relies on the
collinear factorization of the cross section into non-perturbative collinear
parton distribution functions (PDFs), hard partonic scattering cross sections
calculable in perturbation theory, and non-perturbative collinear fragmentation
functions (FFs) \cite{RevModPhys.67.157}. This paper discusses the measurement
of the cross section for production of charged hadrons ($h^{\pm}$) with high transverse
momenta $p_T$ in muon-nucleon ($\mu$-$N$) scattering at low photon virtualities,
$\mu N\rightarrow \mu' h^{\pm} X$. In the pQCD framework, the lowest-order
contributions to this reaction are (i) photon-gluon fusion (PGF), in which a
virtual photon emitted by the lepton interacts with a gluon inside the nucleon
via the formation of a quark-antiquark pair, $\gamma g\rightarrow q
\overline{q}$, (ii) QCD Compton (QCDC) scattering, in which the photon interacts
with a quark in the nucleon leading to the emission of a hard gluon, $\gamma
q\rightarrow q g$, and (iii) numerous resolved-photon processes. 

The comparison of the calculated cross section to the 
experimentally measured one 
is sensitive to the accuracy with
which the partonic cross section can be calculated in perturbation theory, as
well as to the validity of collinear factorization itself,  
i.e.\ to soft non-perturbative contributions to the production of 
high-$p_T$ hadrons.    
For inclusive
high-$p_T$ hadron or jet production in proton-proton ($p$-$p$) scattering, cross
sections have been measured at FNAL
\cite{Donaldson:1977yz,Adams:1994yu,Apanasevich:2002wt}, CERN
\cite{LloydOwen:1980cp} and BNL
\cite{Adler:2003pb,Abelev:2006uq,Adams:2006nd,Adare:2008qb,Abelev:2009pb,Adare:2011vy}
at center-of-mass system (CMS) energies $\sqrt{s_{pp}}$ from $20$\,GeV
to $200$\,GeV. The comparison of these data to next-to-leading order (NLO) pQCD
calculations \cite{Bourrely:2003bw} shows that while there is good agreement at
$\sqrt{s_{pp}}=200$\,GeV (RHIC), the theory increasingly underestimates
the cross sections with decreasing $\sqrt{s_{pp}}$. The disagreement
reaches up to an order of magnitude at $20$\,GeV.  These discrepancies can be
reconciled by the inclusion of all-order resummations of threshold logarithms
\cite{deFlorian:2005yj}, which are related to soft gluon emissions and are
usually performed up to next-to-leading logarithmic (NLL) accuracy.

The
electromagnetic probe in muon-lepton scattering has the advantage over $p$-$p$
scattering that the kinematics of the reaction is better known since the
momentum and energy transfers to the nucleon can be measured for each event by
analyzing the scattered lepton. In the regime of quasi-real photoproduction,
i.e.\ at low photon virtualities $Q^2$, the cross section for high-$p_T$ hadron
production in lepton-nucleon scattering can be calculated in NLO pQCD via the
Weizs\"acker-Williams formalism \cite{Frixione:1993yw,deFlorian:1999ge}. For
dijet production at HERA at very high photon-nucleon CMS\ energies $142\le
W_{\gamma N} \le 293$\,GeV, the NLO pQCD results agree well with the
experimental data \cite{Chekanov:2007ac}. At the energy of fixed-target
experiments, such a check of the applicability of pQCD to high-$p_T$ particle
production at low $Q^2$ has not been done yet. The cross section for high-$p_T$
hadron production in the scattering of 28\,GeV/$c$ positrons off nucleons has
been published by the HERMES Collaboration \cite{Airapetian:2010ac}.  However,
the measurement hardly exceeds $p_T$ values of 2\,GeV/$c$, which sets rather low
factorization and renormalization scales for pQCD calculations, and a comparison
to NLO pQCD was not attempted. A new measurement of the cross
section for production of unidentified charged hadrons with high $p_T$ in
scattering of 160\,GeV/$c$ muons off nucleons (CMS energy
$\sqrt{s_{\mu N}}=17.4$\,GeV) at the COMPASS experiment
\cite{Abbon:2007pq} at low photon virtualities is described in the 
present paper.    
The cross section
for this kinematic domain has been calculated in 
NLO pQCD \cite{Jager:2005uf,Vogelsang:2012priv}. 
Recently, the all-order resummation of threshold corrections 
up to NLL accuracy 
has been 
included in these calculations \cite{deFlorian:2013taa}. 
 
\section{Experiment and Data Analysis} 
\label{sec:exp}
The hadron-production cross section is measured in bins of $p_T$ and $\eta$ of
widths $\Delta p_T$ and $\Delta \eta$, respectively, and is defined as
\begin{linenomath}
\begin{align}\label{formula_xsect2} 
E\frac{\mbox{d}^{3}\sigma}{\mbox{d}p^3}
= \frac{1}{2\pi p_{T}} \frac{N_{h}}{\Delta p_{T}\cdot \Delta \eta\cdot L\cdot
  \epsilon}\quad ,
\end{align}
\end{linenomath}
where $E$ and $p$ are energy and momentum of the hadron, respectively,
$p_T=p\cdot \sin \theta$ is the transverse momentum of the hadron with respect
to the direction of the virtual photon ($\theta$ is the angle between the
virtual photon and the hadron momenta), and $\eta=-\ln \tan (\theta/2)$ is the
pseudo-rapidity of the hadron, all measured in the laboratory system. The
integrated luminosity is denoted by $L$, $N_h$ is the number of observed hadrons
in a given bin of $p_T$ and $\eta$, and $\epsilon$ is the acceptance-correction
factor, which is determined independently for both hadron charges for each bin
of $p_T$ and $\eta$. This factor corrects the number of observed hadrons for
geometrical acceptance and detection efficiency of the spectrometer as well as
for kinematic smearing. The cross section is defined as a single-inclusive cross
section, i.e.\ several high-$p_T$ hadrons per muon-scattering event are counted
for the hadron yield $N_h$. 

The experimental data were recorded in 2004 with
the COMPASS spectrometer at CERN. In the experiment a naturally-polarized
160\,GeV/$c$ $\mu^{+}$-beam scatters off a polarized, isoscalar target that
consists of granulated $^6$LiD immersed in liquid helium.  The small admixtures
of H, $^3$He, and $^7$Li lead to an excess of neutrons of about 0.1\%.  The
target is arranged in two oppositely polarized 60\,cm long cells. The
unpolarized cross section is obtained by averaging over the target
polarizations. Since the azimuthal angles of the produced hadrons are integrated
over, the cross section does not depend on the beam polarization. The integrated
luminosity is determined via the direct measurement of the rate of beam muons
crossing the target and is found to be equal to 142\,pb$^{-1} \pm
10\%\mbox{(syst.)}$ after correction for the dead times of the veto and data acquisition
systems. As an independent cross check of the luminosity, the structure function
of the nucleon $F_2$ is determined from this data set and compared to the NMC
parametrization of $F_2$ \cite{Arneodo:1995cq} yielding satisfactory agreement
\cite{Hoeppner:2011}. The analysis is based on high-$p_T$ events that were
recorded by the quasi-real photoproduction trigger systems
\cite{Bernet:2005yy}. These triggers are based on the coincidence between the
detection of the scattered muon at low scattering angles and an energy deposit
exceeding about 5\,GeV in one of the two hadronic calorimeters, to suppress
background from muon-electron scattering and radiative elastic or quasi-elastic
muon-scattering events. Events are accepted if the photon virtuality
$Q^2<0.1$\,$($GeV/$c$)$^2$ and if the fractional energy transferred from the
incident muon to the virtual photon is in the range $0.2\leq y \leq 0.8$, where
the acceptance of the trigger systems is largest.  These selections result in
the energy range $7.8\le W_{\gamma N} \le 15.5$\,GeV. The fraction of the
virtual-photon energy transferred to the hadron $h^{\pm}$ is constrained by
$0.2\leq z \leq 0.8$. Moreover, hadrons are required to have momenta $p\geq
15$\,GeV/$c$ to ensure full trigger efficiency. The angle of the hadron with
respect to the direction of the virtual photon has to be in the range $10\leq
\theta \leq120$\,mrad, which corresponds to a range of $\mu\text{-}N$ CMS
pseudo-rapidities $2.4\geq \eta_{\text{CMS}} \geq -0.1$. In addition to these
\emph{kinematic} criteria, the selection of reconstructed hadrons is subject to
several \emph{geometrical} cuts: the positions of the muon-scattering vertices
are limited to the fiducial target volume, the hadron tracks must not cross the
solenoid magnet of the polarized target, and the hadron tracks must hit one of
the two hadronic calorimeters, excluding 3\,cm wide margins around the edges
(for full trigger efficiency). 

The acceptance correction factors of Eq.\
(\ref{formula_xsect2}) are determined with a Monte-Carlo (MC) simulation of
$\mu$-$N$ scattering in the COMPASS experiment.  Events are generated with
PYTHIA6 \cite{Sjostrand:2000wi}, the response of the spectrometer is simulated
with a GEANT3-based program \cite{Brun:1993}, and the data are reconstructed
with the same software as the experimental data \cite{Abbon:2007pq}.  The
acceptance factor for the bin $p_T\in[p_{T,1},p_{T,2}]$ is defined as
\begin{linenomath}
\begin{align}\label{formula_acc} \epsilon
=
\frac{N^{\text{rec}}(p_T^{\text{rec}}\in[p_{T,1},p_{T,2}])}{N^{\text{gen}}(p_T^{\text{gen}}\in[p_{T,1},p_{T,2}])}\quad
, \end{align}
\end{linenomath}
where $N^{\text{rec}}$ is the number of reconstructed hadrons in the bin of
reconstructed transverse momentum $p_T^{\text{rec}}$, and $N^{\text{gen}}$ is
the number of generated hadrons in the MC sample in the bin of generated
transverse momentum $p_T^{\text{gen}}$. While both $N^{\text{rec}}$ and
$N^{\text{gen}}$ are subject to the above-listed \emph{kinematic} selection
criteria, the \emph{geometrical} cuts are only applied to $N^{\text{rec}}$ so
that the loss of hadrons due to these cuts is accounted for by the acceptance
correction.

Hadrons that are created at the $\mu$-$N$ vertex constitute the
signal of the measurement and have to be separated from background hadrons,
which are created in secondary interactions of other hadrons in the target
material. This separation is performed by the vertex-reconstruction algorithm,
which is however impaired by the fact that the angle between the incoming and
outgoing muon tracks is very small at low $Q^2$. The background contamination
can not be estimated directly from the MC data, because simulations with the two
hadron-shower models available in GEANT3 (GHEISHA and FLUKA) give inconsistent
results. Hence the background contribution is determined in each $p_T$ bin from
the experimental data by fitting the shape of the distribution of position
differences between two-particle vertices formed by the incoming muon track and
the outgoing muon track on the one hand, and the incoming muon track and the
outgoing hadron track on the other hand \cite{hoeppner:2012}.  The distribution
for signal hadrons, originating from the same interaction as the outgoing muon
track, has a symmetric shape, while for background hadrons there is a
characteristic asymmetric shape.  The results of these fits show that the
background contribution to the experimental data is consistent with
zero. However, cross checks with both MC hadron-shower models 
indicate that the background
contribution can be systematically underestimated by 6\% using this method. In
addition, the described procedure is statistically limited for the highest $p_T$
bins because there are too few entries in the vertex-difference distributions to
exclude a non-zero background contribution with high statistical accuracy. For
the four highest $p_T$ bins, the background level $p_{\text{excl}}$ at which a
non-zero background contribution can be excluded at 90\% confidence level is
greater than 6\%.  Therefore, the possible contribution of residual background
to the hadron yield is conservatively estimated to be $2\times 6\%$ for the six
lowest $p_T$ bins and $p_{\text{excl}}+6\%$ for the four highest $p_T$
bins. These values are used as systematic uncertainties of the acceptance
factors. 
 
A second contribution to the systematic uncertainties of the
acceptance factors arises from the fact that they are determined in a
one-dimensional way, i.e.\ by integrating over all kinematic variables other
than $p_T$.  The resulting uncertainty is quantified by calculating the
acceptance correction binned in two variables, i.e. $p_T$ and one of the
variables $Q^2$, $y$, $x_{\text{Bj}}$ (Bjorken scaling variable), $W_{\gamma
N}$, $z$, $\theta$. A comparison of the cross section calculated in two
variables, summed up over the second variable, with the one-dimensional result
yields deviations below 3\%.  This uncertainty is added in quadrature to the
uncertainties from background contamination, resulting in the following
definition of the upper ($\epsilon_u$) and lower ($\epsilon_d$) limits of the
systematic uncertainty band of the acceptance factors
\begin{linenomath}
\begin{align*}\label{formula_acc_syst} 
 \epsilon_u &=
 \epsilon\cdot \left( 1+\sqrt{0.03^2+(0.06+\max{(0.06,p_{\text{excl}})})^2}
 \right)\quad, \\
 \epsilon_d &= \epsilon \cdot(1-0.03)\quad .
 \end{align*}
\end{linenomath}
Another systematic uncertainty of the cross section is the 10\% normalization
uncertainty from the luminosity determination.
A dependence of the $p_T$ distribution of hadrons on the nuclear 
medium 
has not been observed at COMPASS energies \cite{Adolph:2012vj}. 

\section{Results}
\label{sec:results}
\begin{figure}[t]
\begin{minipage}[t]{0.48\textwidth}
\includegraphics[width=\textwidth]{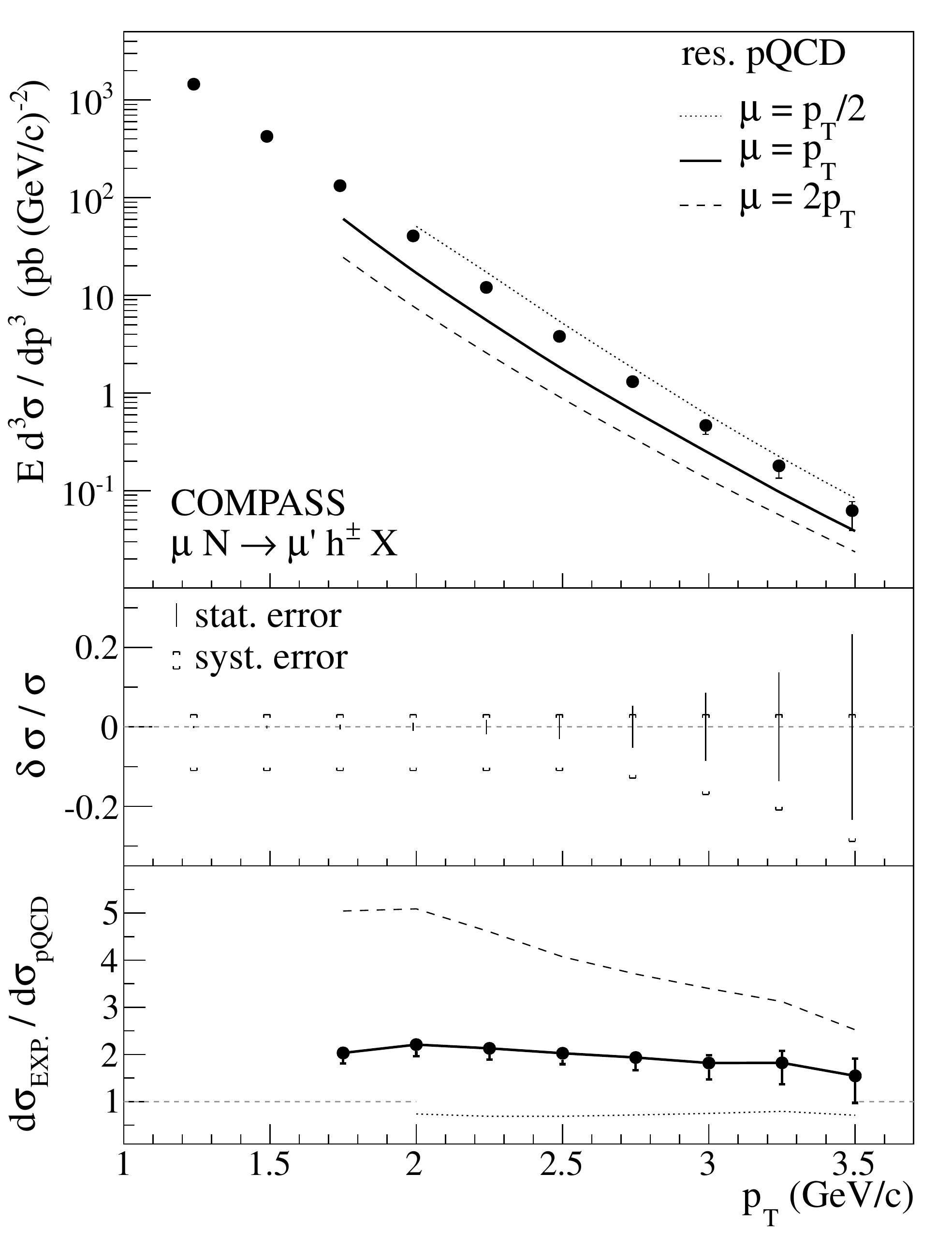}
\caption{Upper panel: differential cross section in bins of $p_T$ 
  for high-$p_T$  
  hadron production in  
  $\mu$-$N$ scattering (data points),  
  compared to
  the resummed pQCD calculation \cite{deFlorian:2013taa} (lines).  
  The other kinematic variables have been integrated over. 
  Middle panel: relative statistical and systematic
  uncertainties of the measurement. Lower panel: ratio of
  the measured over calculated cross sections.}
\label{fig:XS}
\end{minipage}
\hspace{0.035\textwidth}
\begin{minipage}[t]{0.48\textwidth}
\includegraphics[width=\textwidth]{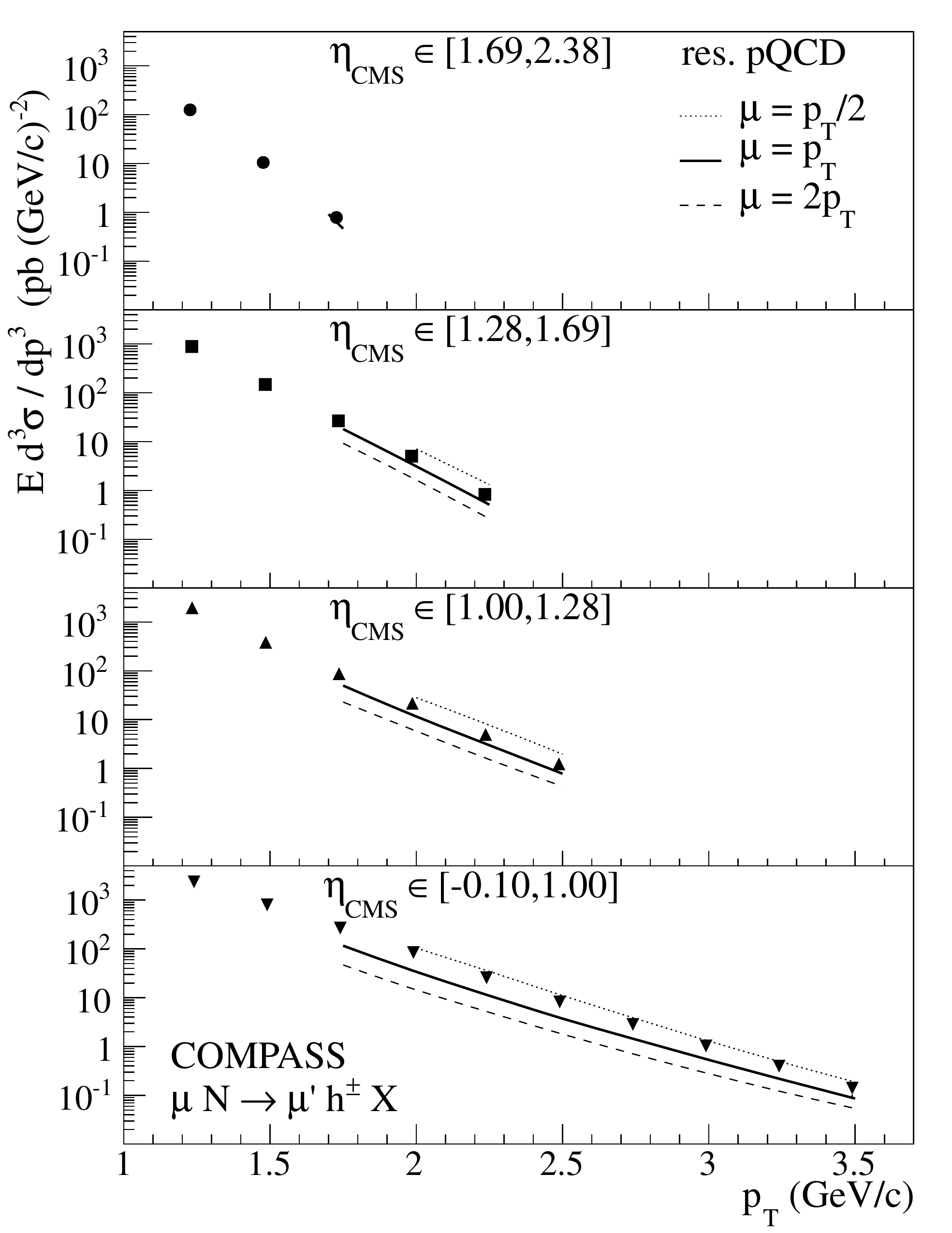}
\caption{$p_T$-differential cross section for four bins of $\eta_{\text{CMS}}$ (data points),  
  compared to
  the resummed pQCD calculation \cite{deFlorian:2013taa} (lines).}
\label{figEta}
\end{minipage}
\end{figure}

The differential cross section in bins of $p_T$ for the production of 
charged high-$p_T$ hadrons in $\mu$-$N$ 
scattering at $Q^2<0.1$\,(GeV/$c$)$^2$ and $\sqrt{s_{\mu N}}=17.4$\,GeV   
is presented in Fig.~\ref{fig:XS} and 
listed in Table~\ref{tab:xs}. 
The errors 
in the upper and lower panels are the quadratic sums of statistical and systematic
uncertainties. The normalization uncertainty of 10\% from
the luminosity measurement is not shown.  
The cross section values are not corrected for
QED radiative effects. These have been estimated to be smaller than 
$5\%$ in the kinematic region of the underlying data sample  
\cite{Akushevich:1999,Afanasev:2013priv}. 
\begin{table*}[tbp]
\centering
\caption{Measured cross section for high-$p_T$ hadron production in $\mu$-$N$
  scattering at $\sqrt{s_{\mu N}}=17.4$\,GeV. The cross section is
  integrated over the full kinematic range defined in the text. The columns
  show: (1) $p_T$ range of the bin; (2) $p_T$ value of data point in
  Fig.~\ref{fig:XS}; (3) differential cross section summed over hadron charges
  (please note that there is an additional 10\% normalization uncertainty from
  luminosity); and (4) charge ratio of the cross section.}
\label{tab:xs}
\small
\begin{tabular}{ll r@{$\pm$}l@{\,(stat.)~}l@{\,(syst.)}l l@{$\pm$}l@{\,(stat.)}}\hline\hline
$[p_{T,1},p_{T,2}]$\,(GeV/$c$) & $\langle p_{T}\rangle_{\text{lw}}$\,(GeV/$c$) & 
\multicolumn{4}{c}{$\frac{\text{d}\sigma}{\text{d}p_T}=\frac{1}{p_{T,2}-p_{T,1}} \int_{p_{T,1}}^{p_{T,2}}\frac{\text{d}\sigma}{\text{d}p_T}\text{d}p_T$\,(pb(GeV/$c$)$^{-1}$)} & \multicolumn{2}{c}{$\frac{\text{d}\sigma}{\text{d}p_T}(\text{h}^-) / \frac{\text{d}\sigma}{\text{d}p_T}(\text{h}^+)$}\\ \hline
$[1.125,1.375]$ & $1.239$ & $[2.810$ & $0.006$ & $_{-0.310}^{+0.087}$ & $]\cdot 10^{4}$ & $0.874$ & $0.004$ \\ 
$[1.375,1.625]$ & $1.489$ & $[9.87$  & $0.04$  & $_{-1.09}^{+0.31}$  & $]\cdot 10^{3}$ & $0.864$ & $0.007$ \\ 
$[1.625,1.875]$ & $1.739$ & $3603$   & $23$    & $_{-397}^{+112}$   &                 & $0.850$ & $0.011$ \\ 
$[1.875,2.125]$ & $1.989$ & $1261$   & $14$    & $_{-139}^{+40}$    &                 & $0.829$ & $0.018$ \\ 
$[2.125,2.375]$ & $2.239$ & $421$    & $8$     & $_{-47}^{+14}$     &                 & $0.800$ & $0.030$ \\ 
$[2.375,2.625]$ & $2.489$ & $148$    & $5$     & $_{-17}^{+5}$     &                  & $0.85$  & $0.06$  \\ 
$[2.625,2.875]$ & $2.739$ & $55.9$   & $3.0$   & $_{-7.3}^{+1.8}$   &                  & $0.83$  & $0.09$  \\ 
$[2.875,3.125]$ & $2.989$ & $21.7$   & $1.9$   & $_{-3.7}^{+0.7}$   &                  & $0.78$  & $0.14$  \\ 
$[3.125,3.375]$ & $3.239$ & $9.08$   & $1.25$  & $_{-1.90}^{+0.29}$ &                  & $0.80$  & $0.23$  \\ 
$[3.375,3.625]$ & $3.490$ & $3.40$   & $0.80$  & $_{-0.98}^{+0.11}$ &                  & $1.0$   & $0.5$ \\ \hline\hline
 \end{tabular}\\
\end{table*}
The discrete $p_T$ values, at which the cross section
values from the binned analysis of Eq.\ (\ref{formula_xsect2}) are drawn, are
calculated using the method of Lafferty \& Wyatt \cite{Lafferty:1994cj} and are
denoted by $\langle p_T\rangle_{\text{lw}}$ in Table~\ref{tab:xs}. 
The cross
section drops by about four orders of magnitude over the measured $p_T$
range. The only apparent deviation from an exponential shape is a slight
hardening of the spectrum at about $p_T=2.5$\,GeV/$c$. In Fig.~\ref{fig:XS}, the
data are compared to an NLO pQCD calculation. 
The method of the calculation is first described in  
Ref.~\cite{Jager:2005uf}, and  
has been updated \cite{Vogelsang:2012priv} to implement the 
kinematic selections presented in Section~\ref{sec:exp}  
and the DSS FFs \cite{deFlorian:2007aj} for unidentified
charged hadrons. Recently, the resummation of large logarithmic 
thresholds to all orders \cite{deFlorian:2013taa} has been included.  
The three curves correspond to different choices of the
renormalization ($\mu_r$) and factorization ($\mu_f$) scales in the pQCD
calculation.  The standard choice for the scales in pQCD is
$\mu=\mu_r=\mu_f=p_T$ and the scale uncertainty is estimated by varying the
scale in the range $p_T/2\le\mu\le 2 p_T$. 
The theoretical values are given only for 
$p_T \ge 1.75$\,GeV/$c$ in order to ensure the applicability of perturbative 
methods. 
At the standard scale $\mu=p_T$, the resummed result underestimates 
the experimental
cross section by a factor of about two, but follows the shape of the
differential cross section remarkably well, 
as can be seen in Fig.~\ref{fig:XS} (bottom panel), which shows the ratio 
of the measured over the calculated cross sections.   
Analogous to $p$-$p$ scattering at low CMS energies 
\cite{Bourrely:2003bw,deFlorian:2005yj}, the 
all-order resummation of threshold logarithms is found to significantly 
reduce the normalization discrepancy compared to the fixed-order 
NLO result \cite{Vogelsang:2012priv}, which underestimated the experimental 
cross section by a factor of three to four.  
The large scale uncertainty of the theoretical cross section, however, 
shows that higher-order 
contributions 
are likely to be significant in the pQCD
framework.

In Fig.~\ref{figEta}, the $p_T$ dependence of
the experimental cross section is presented in bins of $\eta_{\text{CMS}}$,
together with the comparison to the resummed pQCD results. 
The errors are the quadratic sums 
of statistical
and systematic uncertainties, and are smaller than the symbols, 
except for the highest $p_T$ values.  
As in Fig.~\ref{fig:XS}, the normalization uncertainty of 10\% from the 
luminosity measurement is not shown.
The steeper $p_T$ slopes of the cross 
section at forward rapidities as compared to central rapidity are well described
by the pQCD curves. 
The normalization difference between the theoretical
calculation ($\mu=p_T$) and the experimental values 
shows a slight increase towards smaller pseudo-rapidities. 

\begin{figure*}[tbp]
  \centering
  \includegraphics[width=0.75\textwidth]{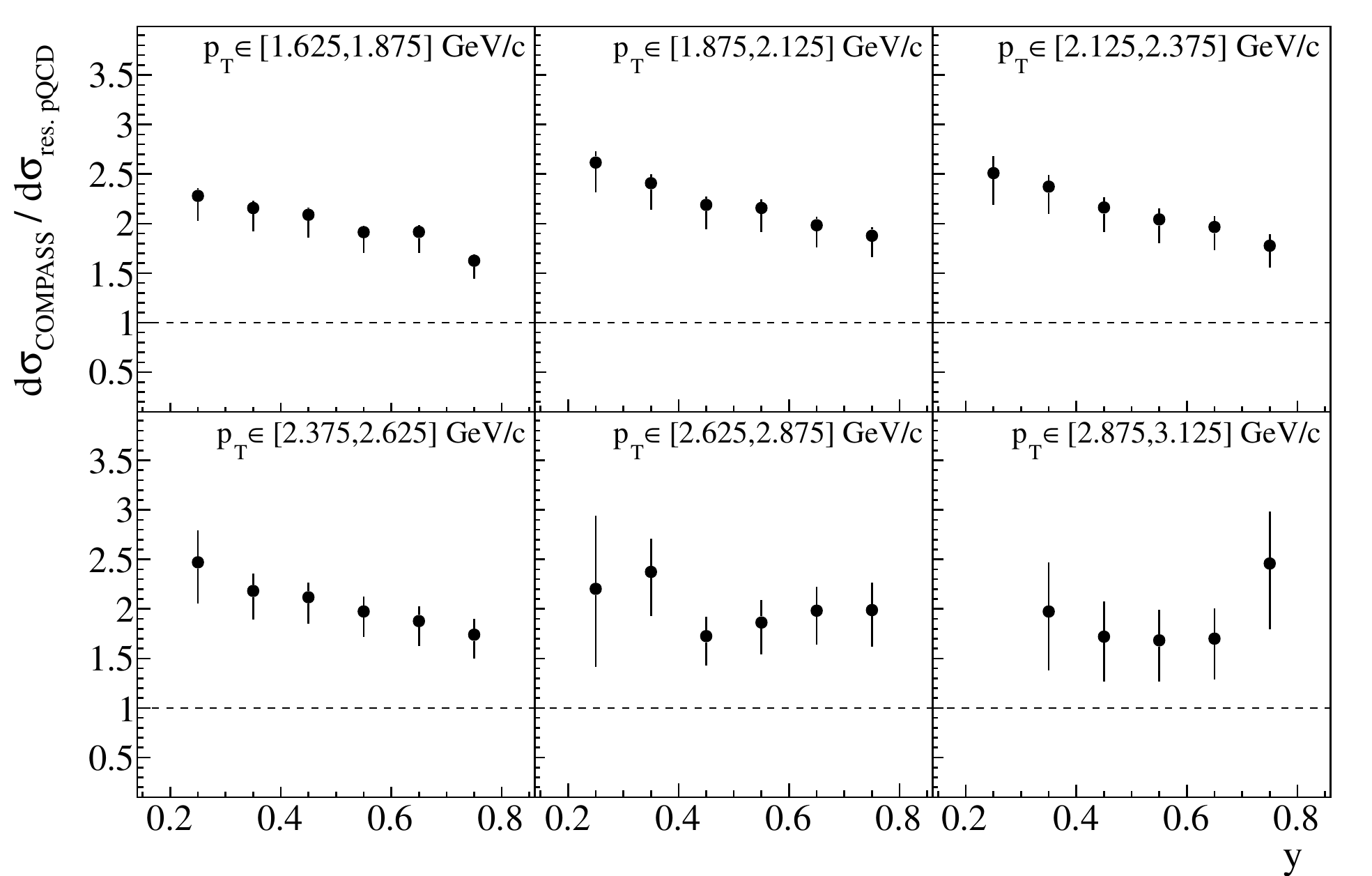}
  \caption{Ratio of $y$-dependent cross section measured by COMPASS and calculated 
    in pQCD \cite{deFlorian:2013taa}, including the 
    resummation of threshold logarithms ($\mu=p_T$), in bins of $p_T$. 
    The errors are the quadratic sums 
    of statistical
    and systematic uncertainties.}  
  \label{figY}
\end{figure*}

In order to judge whether hadron production at the COMPASS kinematics is 
correctly described by pQCD, it is interesting to investigate whether the 
cross section ratio between theory and experiment depends on 
the virtual photon energy fraction $y$. At fixed transverse momentum $p_T$, the phase space 
for the production of additional partons decreases with decreasing $y$. Corrections 
due to the emission of soft gluons are therefore expected to be larger for smaller $y$. 
Figure~\ref{figY} compares the ratio of the COMPASS measurement and the resummed pQCD 
calculation at $\mu=p_T$ of the double differential cross section 
$\mathrm{d}^2\sigma/(\mathrm{d}p_T\mathrm{d}y)$ in six $p_T$ bins, 
integrated over the $p_T$ bin widths:
\begin{equation}
  \frac{1}{0.1}\int_{y-0.05}^{y+0.05}\mathrm{d}y'\int_{p_{T,a}}^{p_{T,b}}
  \frac{\mathrm{d}^2\sigma}{\mathrm{d}p_T\mathrm{d}y'}\mathrm{d}p_T \quad. \nonumber 
\end{equation}
The fact that the cross section ratio depends only weakly on $y$ 
indicates that the resummation 
procedure correctly 
includes the contribution of soft gluon emission to the cross section.  

\begin{figure}[tbp]
  \begin{center}
    \includegraphics[width=0.55\columnwidth]{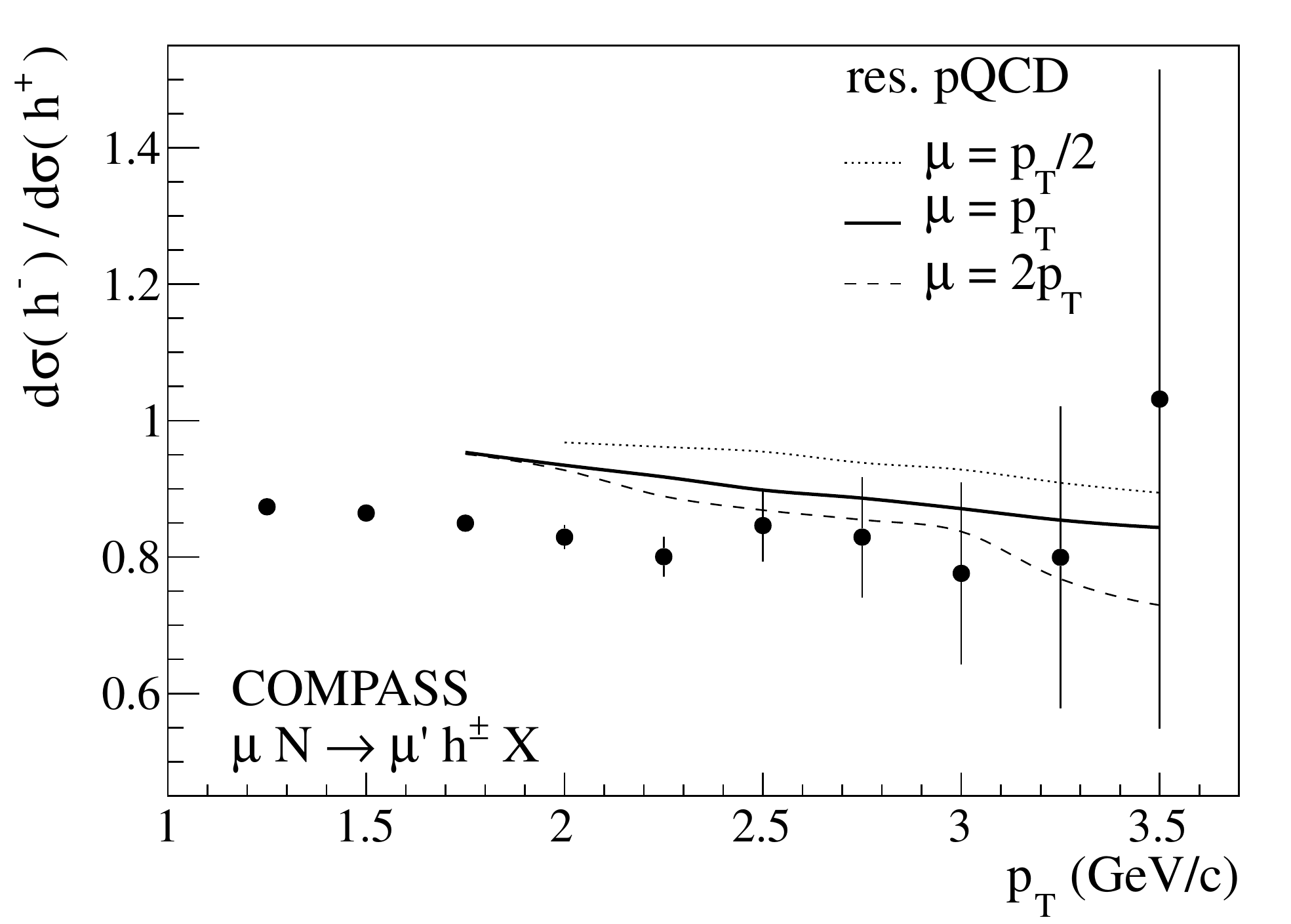}
    \caption{Ratio of cross sections for production of $h^{-}$ over $h^{+}$ 
      as a function of $p_T$. The
      data are compared to the resummed pQCD calculation \cite{deFlorian:2013taa}. 
    }
    \label{figRatio}
  \end{center}
\end{figure}

The ratio of the cross sections for the production of
negatively over positively charged hadrons (charge ratio), displayed in 
Fig.~\ref{figRatio}  
as a function of $p_T$, 
is found to be significantly smaller than unity, showing that the production of 
positive hadrons is preferred. 
No strong $p_T$ dependence is observed within the statistical accuracy of the
measurement. 
It is worth to note that most of the systematic uncertainties 
as well as the normalization uncertainty 
are expected to cancel out in the charge ratio. 
The ratio is sensitive to the contributions of the
different partonic processes to the cross section. The QCDC process can lead to
an excess of positively charged hadrons because the electromagnetic coupling to
$u$ quarks is four times larger than to $d$ quarks, and $u$ quarks are more
likely to produce positively charged mesons. The PGF process, on the other hand,
is not expected to result in a charge asymmetry, assuming independent quark
fragmentation. The resummed pQCD calculation, 
also shown in Fig.~\ref{figRatio},
features a charge ratio of about unity for the lowest $p_T$ values, 
in disagreement with the data,  
and a clear decrease
with increasing $p_T$. It should be noted, however, that the 
scale uncertainty bands were obtained simply by dividing the calculated $h^-$ and $h^+$  
cross sections for a given scale, and thus 
may underestimate the true scale uncertainty \cite{deFlorian:2013taa}. 

\section{Conclusions}
\label{sec:conclusions}
In summary, the
single-inclusive cross section for charged-hadron production in $\mu$-$N$
scattering at $\sqrt{s_{\mu N}}=17.4$\,GeV was measured for photon
virtualities $Q^2<0.1$\,(GeV/$c$)$^2$ in the $\eta_{\text{CMS}}$ interval
between $-0.1$ and $2.4$ and for transverse hadron momenta up to
$3.6\,$GeV/$c$. The measured $p_T$-differential cross section 
is compared with pQCD calculations. 
Without the all-order resummation of threshold logarithms, 
the pQCD calculation at
NLO appears to be insufficient to fully describe high-$p_T$ hadron production in
$\mu$-$N$ scattering at low $Q^2$ in the kinematic domain of COMPASS. The
resummation helps to resolve this
discrepancy at least partly. 
At a renormalization and factorization scale corresponding to $p_T$, 
the calculation reproduces the shape of the measured cross section 
over the full rapidity 
range, but  
underestimates the experimental cross section by about a 
factor of two, independent of $p_T$.
Due to the low values of $p_T$ and $\sqrt{s_{\mu N}}$, however, 
the theory still shows a rather large scale dependence, with an uncertainty 
band which overlaps with the experimental data.   
The ratio of the measured cross section and the calculated one is found to 
depend only weakly on the photon fractional energy $y$, indicating that 
the resummation procedure correctly takes into account corrections due to 
the emission of soft gluons. 
The ratio of cross sections for the production of negative over positive 
hadrons is 
found to be always smaller than unity in the full $p_T$ range under 
investigation, 
with no strong dependence on $p_T$. This is in contrast to the theory, 
which shows a  
ratio close to unity for low $p_T$ values. 

As a next step, the pQCD framework will be employed to constrain the 
polarization of 
gluons in the nucleon \cite{Jager:2005uf}, using the double-spin asymmetry of
single high-$p_T$ hadron production at low $Q^2$ extracted from the full COMPASS
muon-scattering data set. This approach is complementary to previous
measurements of the gluon polarization by COMPASS using spin-dependent, high-$p_T$
hadron-pair production \cite{Ageev:2005pq,Adolph:2012}, which employ the MC
generators PYTHIA and LEPTO \cite{Ingelman:1996mq}, respectively, to quantify
the contribution of PGF to the cross section.

\section*{Acknowledgments}
We thank W.~Vogelsang and 
M.~Pfeuffer for
many useful discussions and for providing the pQCD calculations, and A.~Afanasev for 
estimating the QED radiative corrections. 
We acknowledge the support of the CERN management and staff, as well as the skills
and efforts of the technicians of the collaborating institutions. Special thanks
go to V.\ Anosov and V.\ Pesaro for their technical support during the
installation and the running of this experiment. 
This work was made possible
thanks to the financial support of our funding agencies.

\bibliographystyle{utcaps2}
\bibliography{hipt_cernstyle}

\end{document}